# Comparison of Predictions of the PC-SAFT Equation of State and Molecular Simulations for the Metastable Region of Binary Mixtures

Alexander Keller[a], Kai Langenbach[1,a], Hans Hasse[a]

[a] *University of Kaiserslautern, Laboratory of Engineering Thermodynamics, Erwin-Schrödinger-Str. 44, D-67663 Kaiserslautern, Germany*

**Abstract**

The metastable region of binary mixtures is examined with the PC-SAFT equation of state (EOS) and compared to molecular simulations. The studied mixtures are Methane + Ethane, Methanol + Ethanol, Carbon Dioxide + Toluene, Carbon Dioxide + Hydrogen Chloride and Hydrogen Chloride + Toluene. In order to calculate the spinodal, the second partial derivatives with respect to the components molarities are analytically determined for the PC-SAFT EOS. Thermal properties as well as the spinodal points determined using PC-SAFT and molecular simulation are compared. For the studied metastable gas phases the results obtained with PC-SAFT and the molecular simulations agree well. For liquid phases the metastable region calculated with the PC-SAFT EOS is typically broader than that obtained from the molecular simulations. No unphysical results were obtained with PC-SAFT for metastable states. Hence, PC-SAFT EOS is found to be principally suitable for predicting metastable states and spinodals of mixtures.

---

[1]Corresponding author; kai.langenbach@mv.uni-kl.de

# 1. Introduction

Metastable states are omnipresent. For example boiling requires superheating, condensing requires supercooling. Effects like retardation of boiling or cavitation are due to metastablility. However, experimental studies of metastable states are difficult. Using molecular simulations, it is possible to obtain reliable information on the metastable region. Molecular simulations rely only on Newtonian mechanics and statistical physics, which hold both for the stable and metastable region. However, the potentials which are used are simplified pictures of reality and yield good results only due to their parameterization [1]. These parameters are state-independent, and it is known that extrapolations based on such potentials are often successful even far away from state points, which were considered in the parameterization. The same holds for properties which were not considered in the parameterization. Moreover, even bold extrapolations based on molecular models are generally well-behaved, i.e. do not yield qualitatively wrong results [2,3]. Hence, a molecular model, which describes properties of stable state points well, can be expected to yield reliable predictions in the metastable region. Also equations of state (EOS) can be used. In contrast to molecular simulation the theories on which EOS are based generally contain semi-empirical approximations. Hence, it is necessary to check their validity in the metastable region. This is done here for an EOS from the statistical associating fluid theory (SAFT) family [4,5].

Consequences of the semi-empirical nature of the EOSs from the SAFT-family have attracted some attention in the literature. For example Privat et al. [6] discussed if the perturbed-chain statistical associating fluid theory (PC-SAFT) EOS generates unphysical results, when it is applied to pure substances at high pressures and high temperatures. Polishuk [7] found multiple vapor-liquid critical points as well as crossing isotherms at very high pressures for several pure compounds using different EOS from the SAFT-family. Yelash et al. [8] also found multiple critical points, including liquid-liquid critical points, in pure compounds consisting of large molecules at very low temperatures for several SAFT versions. Up to now, metastable states were not in the focus of such investigations. In particular, no systematic investigations on metastable states in mixtures are available regarding vapor-liquid equilibria (VLE). Spinodals in liquid-liquid equilibria (LLE) have been calculated with the simplified PC-SAFT EOS by van Kouskoumvekaki et al. [9].

Therefore in the present work, a systematic comparison is made between results from the PC-SAFT EOS [10,11] and molecular simulations, in order to assess the validity of PC-SAFT in the metastable region of mixtures. Binary mixtures are chosen such that they include mixtures of different types: unpolar + unpolar, H-bonding + unpolar, and H-bonding + H-bonding compounds. Size disparity is also investigated. It is examined if these differences in polarity and size have an impact on the quality of the prediction.

To calculate the limits of the metastable region, second derivatives of the free energy with respect to the component molarities are needed. In the literature, usually numerical derivatives are used when applying the PC-SAFT EOS for phase stability problems, e.g. [12,13]. In the present work, the second partial derivatives with respect to the components molarities are calculated analytically for the PC-SAFT EOS. The expressions are given in the Supplementary Information.

For all studied mixtures, thermodynamic states at different partial densities are simulated. The models for the molecular simulations are taken from previous work of our group [14–18]. Models for the PC-SAFT EOS are fit to mixture VLE predictions of molecular simulations or taken from the literature if already suita-ble [10,11]. The VLE predictions for the three binary subsystems of the system Carbon Dioxide//Hydrogen Chloride//Toluene are taken from previous work of our group [19].

**2. PC-SAFT Equation of State**

The PC-SAFT EOS [10, 11] is written in terms of the molar free energy $a$ as a sum of the following contributions:

$$a = a_{id} + a_{hc} + a_{disp} + a_{assoc}, \qquad (1)$$

Where $a_{id}$ is the molar free energy of the ideal gas, $a_{hc}$ is the contribution due to a reference hard-chain fluid [20,21], $a_{disp}$ is the contribution due to dispersion interactions between the chains [10,11,22,23] and $a_{assoc}$ is the contribution due to association [11]. The PC-SAFT version employed in the present work is taken from [24]. Other versions of PC-SAFT use a slightly different approximation for the association strength. More information is given in the Supplementary Information.

The PC-SAFT parameters for the pure substances used in this work are given in Table 1. Here, $A$, $P$ and $E$ refer to associating acidic, protonic and electronegative sites as in [25]. In the dispersion term for mixtures the binary interaction parameter $k_{ij}$ is used:

$$\varepsilon_{ij} = (1 - k_{ij})\sqrt{\varepsilon_i \cdot \varepsilon_j}, \qquad (2)$$

The $k_{ij}$ values are taken from Werth et al. [19] or set to zero. The values are listed in Table 2.

[Table 1 about here.]

[Table 2 about here.]

From the free energy $A(T,V,\underline{n})$ the spinodal can be determined. Analytical expressions for the second partial derivatives of the PC-SAFT EOS in terms of the free energy $A$ with respect to the components' partial densities $\rho_i$, which are needed for calculating the spinodals, are presented in the Supplementary Information for the first time. For these derivatives, the volume of the mixture $V$ and the temperature $T$ are held constant. According to Heidemann and Khalil, a spinodal composition for a system with $N$ components is found, when the determinant of the Hessian matrix of $A(\underline{n})$ equals zero [26]:

$$\det \begin{pmatrix} \dfrac{\partial^2 A(n_1,...,n_N)}{\partial n_1 \partial n_1} & \dfrac{\partial^2 A(n_1,...,n_N)}{\partial n_1 \partial n_2} & \cdots & \dfrac{\partial^2 A(n_1,...,n_N)}{\partial n_1 \partial n_N} \\ \dfrac{\partial^2 A(n_1,...,n_N)}{\partial n_2 \partial n_1} & \dfrac{\partial^2 A(n_1,...,n_N)}{\partial n_2 \partial n_2} & \cdots & \dfrac{\partial^2 A(n_1,...,n_N)}{\partial n_2 \partial n_N} \\ \vdots & \vdots & \ddots & \vdots \\ \dfrac{\partial^2 A(n_1,...,n_N)}{\partial n_N \partial n_1} & \dfrac{\partial^2 A(n_1,...,n_N)}{\partial n_N \partial n_2} & \cdots & \dfrac{\partial^2 A(n_1,...,n_N)}{\partial n_N \partial n_N} \end{pmatrix}_{V,T} = 0 \ . \quad (3)$$

To evaluate this criterion with the PC-SAFT EOS it is convenient to use $a(\underline{\rho})$ instead of $A(\underline{n})$, where $a$ is the molar free energy and $\underline{\rho}$ is the vector containing the partial densities of the $N$ components, with

$$\rho_i = \frac{n_i}{V} \ . \quad (4)$$

Second derivatives of $A(\underline{n})$ and $a(\underline{\rho})$ are connected by

$$V \frac{\partial^2 A}{\partial n_i \partial n_j} = \frac{\partial^2 (\rho a)}{\partial \rho_i \partial \rho_j} \ . \quad (5)$$

Using $\partial \rho / \partial \rho_i = 1$, this results in

$$\frac{\partial^2 (\rho a)}{\partial \rho_i \partial \rho_j} = \frac{\partial a}{\partial \rho_j} + \frac{\partial a}{\partial \rho_i} + \rho \frac{\partial^2 a}{\partial \rho_i \partial \rho_j} \ . \quad (6)$$

Since the individual terms in the PC-SAFT EOS are additive, the derivatives can be calculated by derivation of these terms, e.g.

$$\frac{\partial^2 a}{\partial \rho_i \partial \rho_j} = \frac{\partial^2 a_{\text{ideal}}}{\partial \rho_i \partial \rho_j} + \frac{\partial^2 a_{\text{hc}}}{\partial \rho_i \partial \rho_j} + \frac{\partial^2 a_{\text{disp}}}{\partial \rho_i \partial \rho_j} + \frac{\partial^2 a_{\text{assoc}}}{\partial \rho_i \partial \rho_j} \ . \quad (7)$$

The results for the individual terms are listed in the Supplementary Information.

## 3. Molecular Models and Simulations

The molecular models of the pure components are taken from previous work of our group [14–18], and are listed in Table 3.

[Table 3 about here.]

The Lorentz-Berthelot mixing rules were used, however with introducing an adjustable, state independent binary interaction parameter $\xi$ in

$$\varepsilon_{ij} = \xi \cdot \sqrt{\varepsilon_i \cdot \varepsilon_j} \ . \tag{8}$$

The values for $\xi$ used in the present work are listed in Table 4. They are taken from previous work of our group [19] or set to 1.

[Table 4 about here.]

All simulations are performed with the *ms*2 simulation code [27]. Metastable states are simulated using the Monte Carlo (MC) method in the canonical (NVT) ensemble. For the MC-runs, the equilibration is performed using 20,000 NVT cycles while the production is performed using 1,000,000 cycles. One MC-cycle consists of 100 displacement moves, 1,000 rotation moves and 1 volume move. A center of mass cut-off radius of 4.5 $\sigma$ is chosen. The number of particles is 1,024.

The VLE of methane and ethane is computed with the Grand Equilibrium method by Vrabec and Hasse [28]. For the liquid phase, molecular dynamics (MD) simulations are performed in the isobaric-isothermal NpT ensemble, using 1,000 particles and Andersen's barostat [29]. In order to calculate the chemical potential, Widom's test particle method [30] is utilized, by inserting up to 5,000 test particles every time step. The cutoff radius is 17.5 Å. The fluid is equilibrated for 50,000 time steps, before sampling during for 3,200,000 time steps. For the corresponding vapor phase, MC simulations are performed in the pseudo-$\mu$VT [28] ensemble. The equilibration is performed using 10,000 NVT cycles and 20,000 NpT cycles, while the production is performed using 100,000 cycles. Each loop consists of $N_{NDF}$ / 3 steps, where $N_{NDF}$ indicates the total number of mechanical degrees of freedom of the system, plus two insertion and deletion attempts. The cutoff radius is equal to half the box length. For the long range correction of both metastable and VLE simulations the reaction field method is used [31].

## 4. Results and Discussion

Numerical data obtained for VLE and homogenous states both with molecular simulations and PC-SAFT are presented in the Supplementary Information. For the present discussion graphical representations are used.

*4.1 Methane + Ethane*

The system Methane + Ethane is studied at 223 K, the results are given in Figure 1. In the upper part of Figure 1 the simulated states are depicted in a $\rho_1$-$\rho_2$-diagram (molarity plane) of the investigated mixture at a constant temperature. In the lower part of Figure 1 the simulated states are depicted in a $\rho_2$-$p$-plot. For the molecular simulations, only the homogenous states are shown in the plot. Simulations in the metastable range that show heterogeneous behavior are neglected, according to one of the following criteria: On the liquid side, the occurrence of a second phase results in an abrupt change of the slope of pressure with respect to the components' molarities. Data points beyond this change of slope are discarded. On the gaseous side, the fluctuations of pressure in a system where a liquid phase occurs are much more pronounced than in a purely gaseous system. Such data points are discarded. Besides the results for the binodal and spinodal obtained with PC-SAFT, results obtained from both PC-SAFT and molecular simulations for VLE and homogenous states with $\rho_{\text{Methane}} = 2$ mol/L and $\rho_{\text{Methane}} = 4$ mol/L are shown. Most of the simulation data lie in the metastable region. The data obtained from both methods agree well, both for VLE as well as for the homogenous states. This is remarkable as the models were developed independently and no adjustment of binary parameters was carried out.

[Figure 1 about here.]

*4.2 Methanol + Ethanol*

In order to investigate the H-bonding term of PC-SAFT, the mixture Methanol + Ethanol is studied. Here, self-association and cross-association occur. Molecular simulations of homogenous states with $x_{\text{Methanol}} = 0.166$ mol/mol are carried out at 353.15 K. In Figure 2 the simulated pressures are compared to PC-SAFT calculations. Most of the simulation data lie in the metastable region. The data obtained from both methods agrees well. Figure 2 looks qualitatively different from Figure 1, because both components are subcritical. Therefore for both spinodal and binodal two lines exist. The same holds true for Figure 4.

[Figure 2 about here.]

*4.3 Carbon Dioxide + Toluene*

The results obtained for the system Carbon Dioxide + Toluene at 353 K are presented in Figure 3. Molecular simulations and PC-SAFT calculations of homogenous states of the gas and the liquid side of the metastable region are carried out for $\rho_{\text{CO2}} = 0.8$ mol/L and $\rho_{\text{CO2}} = 5$ mol/L. The results for the VLE densities from molecular simulations and PC-SAFT agree well. For $\rho_{\text{CO2}} = 0.8$ mol/L the pressure calculated from PC-SAFT is higher than that obtained in the molecular simulations by about 5 %. This, however, holds both for the stable and metastable region.

[Figure 3 about here.]

*4.4 Carbon Dioxide + Hydrogen Chloride*

Hydrogen Chloride shows strong H-bonding, whereas Carbon Dioxide is basically only quadrupolar. The results obtained with molecular simulations and PC-SAFT for that mixture at 290 K are presented in Figure 4. The results for the VLE for both methods agree well. The same holds for the most of the molecular simulations and PC-SAFT calculations of the homogenous states in the metastable region both for the gas and the liquid side, which were obtained for $x_{CO2} = 0.166$ mol/mol. On the liquid side in the region near the spinodal the pressure obtained from the molecular simulations is distinctly larger than that obtained from PC-SAFT.

[Figure 4 about here.]

*4.5 Hydrogen Chloride + Toluene*

The largest disparities in size, as well as H-bonding characteristics of the mixtures investigated here, are found in the system Hydrogen Chloride + Toluene which is studied at 353 K. The results are presented in Figure 5. The results for the VLE obtained from molecular simulations and PC-SAFT agree less well than in the other studied systems. In particular the partial densities of Toluene found in the molecular simulations for saturated liquid states are larger than that obtained from PC-SAFT. Homogenous metastable states are obtained from molecular simulations and PC-SAFT calculations for $\rho_{HCl} = 2$ mol/L and $\rho_{HCl} = 6$ mol/L. Also for the metastable states the numbers obtained for $\rho_{Toluene}$ from the molecular simulations are larger than those from PC-SAFT. The shift seems to be basically induced by the deviations, which are already present in the stable state. We tried to improve the modeling of the mixture with PC-SAFT by adjustment to VLE data from molecular simulation, but were unable to find significant better parameterization. This gives a hint that in PC-SAFT both polarity and H-bonding have to be considered to increase the accuracy in the description of such mixtures. We refrained from using existing polar versions of PC-SAFT e.g. [32] in the present work. The reason for this is that using both the Wertheim term and a polar extension, it is very hard to fit parameters since both of the terms have a similar effect on the free energy. While this does not pose a large problem for pure compounds, in case of mixture the balance between both contributions becomes important. Hence, difficulties can arise from the use of both. Using polar terms, i.e. for dipole-dipole, quadrupole-quadrupole and dipole-quadrupole interaction, might result in improved agreement between molecular simulation and PC-SAFT.

[Figure 5 about here.]

## 5. Conclusions

In the present work the metastable region of the VLE of binary mixtures is examined with the PC-SAFT EOS and compared to molecular simulations. The considered mixtures contain molecules of different types regarding size, polarity and H-bonding.

PC-SAFT predicts physically reasonable results for all studied types of mixtures, i.e. only one van der Waals loop is obtained, and no multiple minima or maxima are observed. Not surprisingly, as the dispersion term of the PC-SAFT EOS is fitted to *n*-alkanes, the results obtained with PC-SAFT for Methane + Ethane agree quantitatively with the molecular simulations. Also in most other studied cases the results obtained for metastable states by PC-SAFT and molecu-lar simulations show good agreement. The deviations are typically of the same order as those obtained for stable states namely VLE.

All in all, the comparison of PC-SAFT to molecular simulations for metastable states for a variety of different mixtures carried out in the present work indicates that PC-SAFT can be reliably used also for quantitative descriptions of metastable states of mixtures. This is especially important for the application of PC-SAFT in connection with the density gradient theory (DGT).

**Acknowledgment**

The authors grateful acknowledge support of this work by the European Research Council in the frame of the ENRICO project.

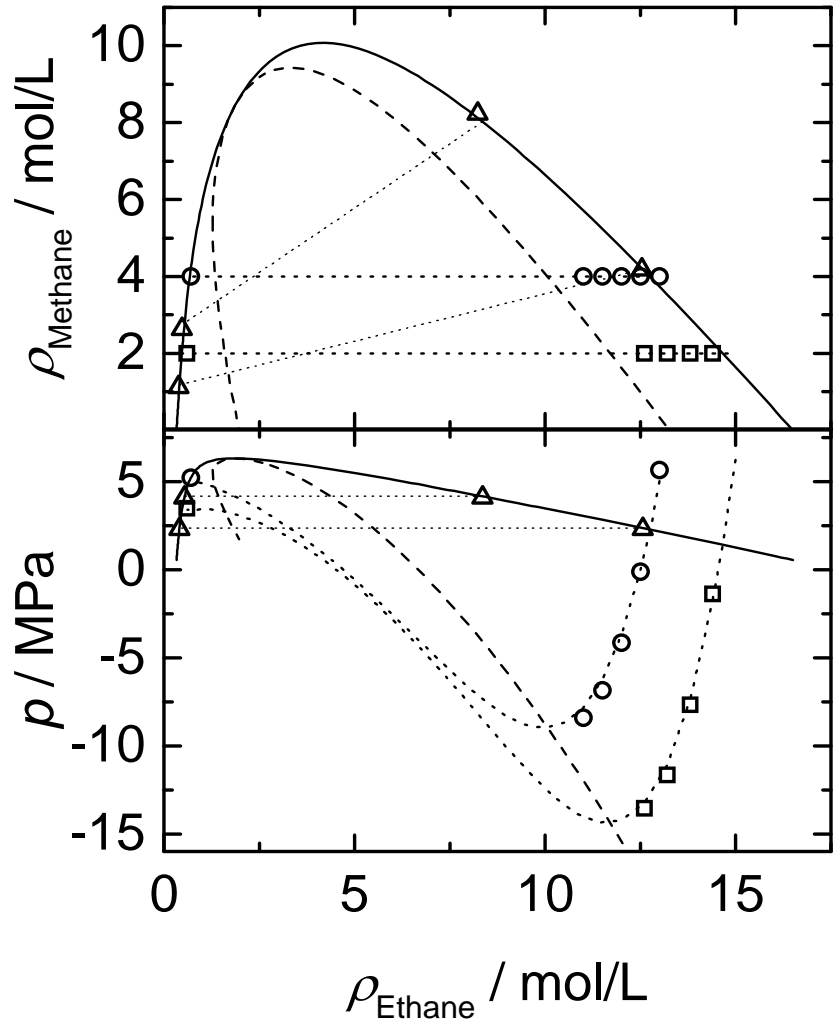

Figure 1: Phase diagram of the mixture Methane + Ethane at 223 K. Binodal (—) and spinodal (---) obtained from PC-SAFT. VLE data obtained from heterogeneous molecular simulations (Δ) and corresponding PC-SAFT calculations (⋯) as well as data for $\rho_{\text{Methane}} = 2$ mol/L and $\rho_{\text{Methane}} = 4$ mol/L obtained from homogenous molecular simulations (□, ○) and from PC-SAFT (⋯).

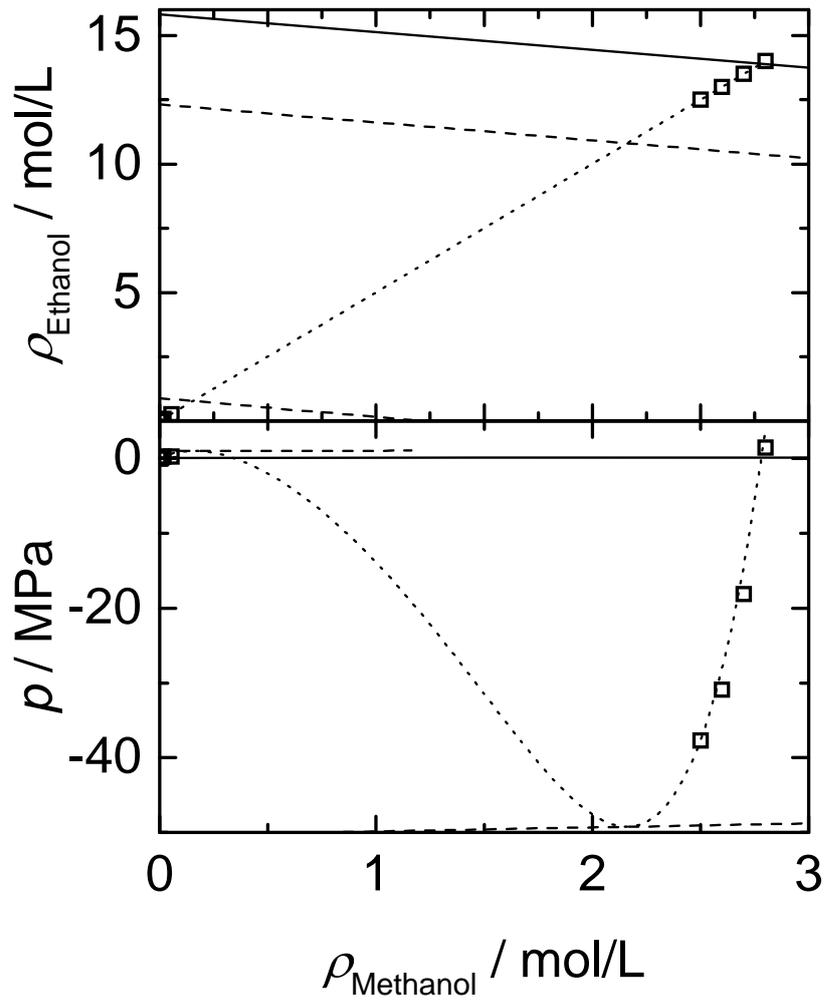

Figure 2: Phase diagram of the mixture Methanol + Ethanol at 223 K. Binodal (—) and spinodal (---) obtained from PC-SAFT. Data for $x_{\text{Methanol}} = 0.1666$ mol/mol L obtained from homogenous molecular simulations (□) and from PC-SAFT (⋯).

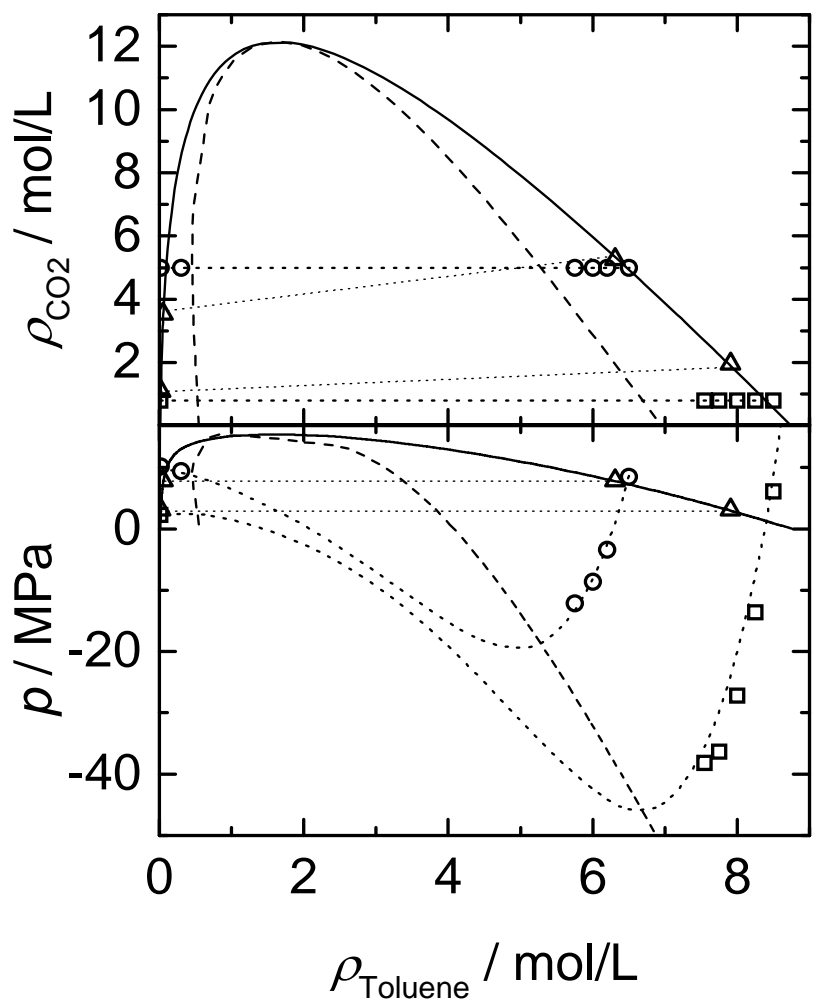

Figure 3: Phase diagram of the mixture $CO_2$ + Toluene at 353 K. Binodal (—) and spinodal (---) obtained from PC-SAFT. VLE data obtained from heterogeneous molecular simulations (△) and corresponding PC-SAFT calculations (⋯) as well as data for $\rho_{CO2} = 0.8$ mol/L and $\rho_{CO2} = 5$ mol/L obtained from homogenous molecular simulations (□, ○) and from PC-SAFT (⋯).

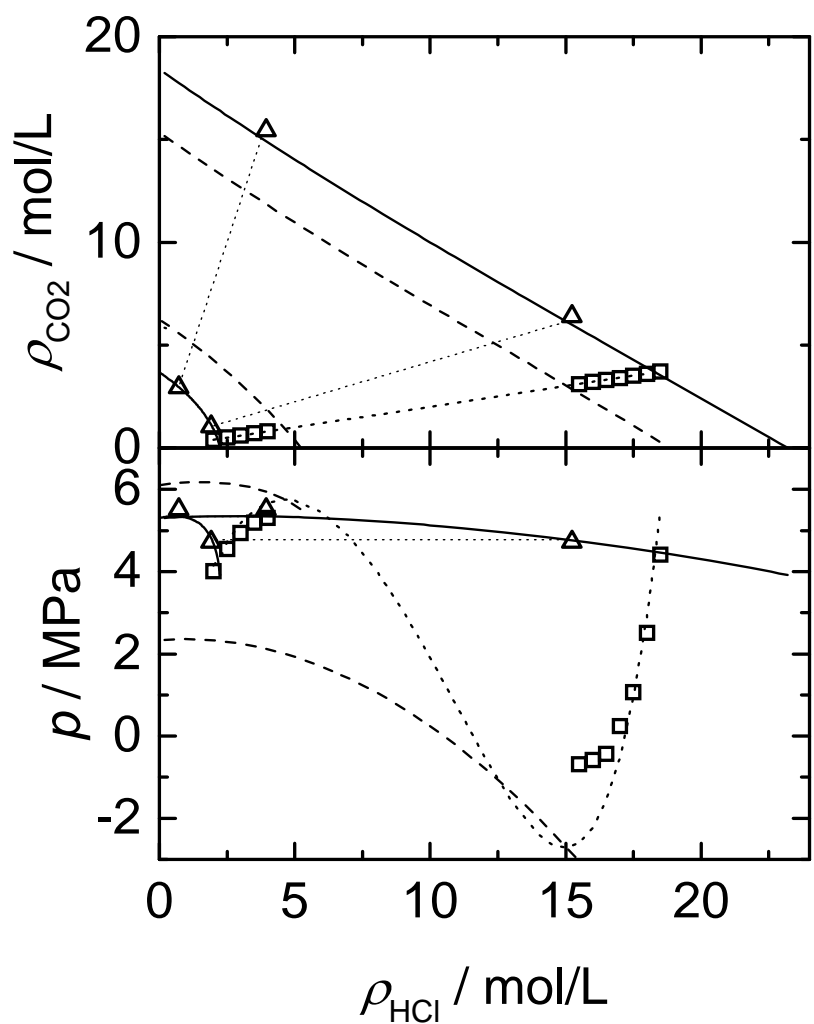

Figure 4: Phase diagram of the mixture $CO_2$ + HCl at 290 K. Binodal (—) and spinodal (---) obtained from PC-SAFT. VLE data obtained from heterogeneous molecular simulations (Δ) and corresponding PC-SAFT calculations (⋯) as well as data for $x_{CO2}$ = 0.166 mol/mol obtained from homogenous molecular simulations (□, ○) and from PC-SAFT (⋯).

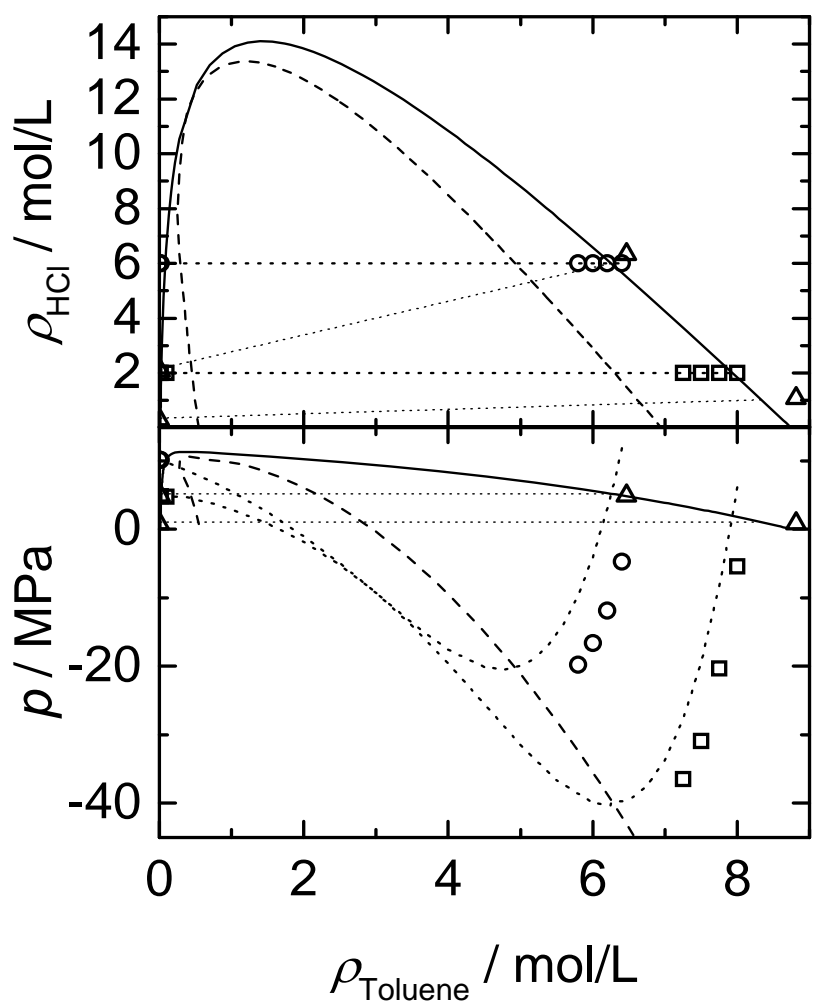

Figure 5: Phase diagram of the mixture HCl + Toluene at 353 K. Binodal (—) and spinodal (---) obtained from PC-SAFT. VLE data obtained from heterogeneous molecular simulations (Δ) and corresponding PC-SAFT calculations (⋯) as well as data for $\rho_{HCl}$ = 2 mol/L and $\rho_{HCl}$ = 6 mol/L obtained from homogenous molecular simulations (□, ○) and from PC-SAFT (⋯).

Table 1: PC-SAFT parameters for the pure substances considered in the present study.

| Fluid | $M$ g/mol | $m$ | $\sigma$ Å | $\varepsilon/k_B$ K | $\varepsilon_{AB}/k_B$ K | $\kappa_{AB}$ | A | P | E | source |
|---|---|---|---|---|---|---|---|---|---|---|
| Carbon dioxide | 44.01 | 2.6779 | 2.5253 | 149.4366 | - | - | - | - | - | this work |
| Ethane | 30.07 | 1.6069 | 3.5206 | 191.4200 | - | - | - | - | - | [10] |
| Ethanol | 46.69 | 2.3827 | 3.1771 | 198.2400 | 2653.4 | 0.032400 | 0 | 1 | 1 | [11] |
| Hydrogen Chloride | 36.46 | 1.6736 | 2.9122 | 202.1071 | 902.9 | 0.000017 | 0 | 1 | 1 | this work |
| Methane | 16.04 | 1.0000 | 3.7039 | 150.0300 | - | - | - | - | - | [10] |
| Methanol | 32.04 | 1.5255 | 3.2300 | 188.9000 | 2899.5 | 0.035176 | 0 | 1 | 1 | [11] |
| Toluene | 92.14 | 2.9283 | 3.6596 | 280.4573 | - | - | - | - | - | this work |

Table 2: Binary PC-SAFT mixing parameter $k_{ij}$ of the studied mixtures.

| Mixture | $k_{ij}$ | source |
|---|---|---|
| Ethane + Methane | 0 | estimated |
| Methanol + Ethanol | 0 | estimated |
| Carbon dioxide + Toluene | 0.115 | this work |
| Carbon dioxide + Hydrogen Chloride | 0.039 | this work |
| Hydrogen Chloride + Toluene | 0.067 | this work |

Table 3: Model parameters of the rigid molecular simulation models.

| Fluid | $x$ Å | $y$ Å | $z$ Å | $\sigma$ Å | $\varepsilon/k$ K | $\theta$ deg | $\varphi$ deg | $q$ e | $\mu$ D | $Q$ DÅ |
|---|---|---|---|---|---|---|---|---|---|---|
| **Carbon dioxide [14]** | | | | | | | | | | |
| C | 0 | 0 | 0 | 2.8137 | 12.3724 | | | | | -4.0739 |
| O (1) | 0 | 0 | -1.2869 | 2.9755 | 100.4931 | | | | | |
| O (2) | 0 | 0 | 1.2869 | 2.9855 | 100.4931 | | | | | |
| **Ethane [16]** | | | | | | | | | | |
| $CH_3$ | 0 | 0 | 1.1881 | 3.4896 | 136.99 | | | | | |
| $CH_3$ | 0 | 0 | 1.1881 | 3.4896 | 136.99 | | | | | |
| Quad | 0 | 0 | 0 | | | | | | | 0.8277 |
| **Ethanol [18]** | | | | | | | | | | |
| $CH_3$ | -1.4707 | -0.3384 | 0 | 3.6072 | 120.15 | | | | | |
| $CH_2$ | $9.2773 \cdot 10^{-2}$ | 0.8833 | 0 | 3.4612 | 86.291 | | | 0.2556 | | |
| OH | 1.1715 | 0.4600 | 0 | 3.1496 | 85.0534 | | | -0.6971 | | |
| Charge | 2.0492 | -0.086 | 0 | | | | | 0.4415 | | |
| **Hydrogen Chloride [15]** | | | | | | | | | | |
| Cl | 0 | 0 | 0 | 3.52 | 179.0 | | | -0.2731 | | |
| H | 0 | 0 | 1.28 | | | | | 0.2731 | | |
| **Methane [16]** | | | | | | | | | | |
| $CH_4$ | 0 | 0 | 0 | 3.7281 | 148.55 | | | | | |
| **Methanol [17]** | | | | | | | | | | |
| $CH_3$ | 0.7660 | 0.0133 | 0 | 3.7543 | 120.5918 | | | 0.2475 | | |
| OH | 0.6565 | -0.0639 | 0 | 3.030 | 87.8790 | | | -0.6787 | | |
| Charge | -1.0050 | 0.8146 | 0 | | | | | 0.4313 | | |
| **Toluene [15]** | | | | | | | | | | |
| $CH_3$ | -2.7164 | -0.0113 | 0 | 3.5864 | 123.4902 | | | | | |
| C | -0.9240 | 0.0205 | 0 | 2.7939 | 10.9358 | 90 | 180 | | 0.4401 | |
| CH (1) | -0.4259 | 0.0152 | 1.5720 | 3.2759 | 100.5188 | 90 | 90 | | | -1.6878 |
| CH (2) | -0.4259 | 0.0152 | -1.5720 | 3.2759 | 100.5188 | 90 | 90 | | | -1.6878 |
| CH (3) | 1.3913 | -0.0028 | 1.5720 | 3.2759 | 100.5188 | 90 | 90 | | | -1.6878 |
| CH (4) | 1.3913 | -0.0028 | -1.5720 | 3.2759 | 100.5188 | 90 | 90 | | | -1.6878 |
| CH (5) | 2.3085 | 0.0137 | 0 | 3.2759 | 100.5188 | 90 | 90 | | | -1.6878 |

Table 4: Binary mixing parameter $\xi$ used in the simulations of the studied mixtures.

| Mixture | $\xi$ | source |
|---|---|---|
| Ethane + Methane | 1.0 | estimated |
| Methanol + Ethanol | 1.0 | estimated |
| Carbon dioxide + Toluene | 0.95 | [19] |
| Carbon dioxide + Hydrogen Chloride | 0.97 | [19] |
| Hydrogen Chloride + Toluene | 0.98 | [19] |